\documentclass[9pt,conference]{IEEEtran}
\usepackage{amsmath,epsfig}
\usepackage{amsthm}
\usepackage{amssymb}
\usepackage{booktabs}
\usepackage{amsthm}
\usepackage{graphicx}
\usepackage{rotating}
\usepackage{floatflt} 
\usepackage{paralist} 
\usepackage{algorithm} 
\usepackage{xcolor}
\usepackage{color}
\usepackage{epstopdf}
\usepackage[normalem]{ulem}
\usepackage{float}
\usepackage{todonotes}
\usepackage{comment}
\usepackage{bm}
\usepackage{mystyle}
\usepackage{subcaption}
\usepackage[font={small}]{caption}
\usepackage[section]{placeins}

\renewcommand{\figurename}{Fig.}
\renewcommand{\tablename}{Tab.}

\usepackage[section]{placeins}

\begin{document}

\title{ Sampling Theory for Graph Signals on Product Graphs}

\twoauthors{Rohan A. Varma, Carnegie Mellon University}{rohanv@andrew.cmu.edu}{Jelena~Kova\v{c}evi\'c, NYU Tandon School of Engineering}{jelenak@nyu.edu}

 \maketitle
\begin{abstract}
In this paper, we extend the sampling theory on graphs by constructing a framework that exploits the structure in product graphs for efficient sampling and recovery of bandlimited graph signals that lie on them. Product graphs are graphs that are composed from smaller~\emph{graph atoms}; we motivate how this model is a flexible and useful way to model richer classes of data that can be multi-modal in nature. Previous works have established a sampling theory on graphs for bandlimited signals. Importantly, the framework achieves significant savings in both sample complexity and computational complexity.
\end{abstract}
\keywords{sampling, graph signal processing, bandlimited, kronecker product}

\section{Introduction}

The task of sampling and recovery is one of the most critical topics in the signal processing community. With the explosive growth of information and communication, signals
are being generated at an unprecedented rate from various sources,
including social networks, citation networks, biological networks,
and physical infrastructure~\cite{Jackson:08,Newman:10}. Unlike time-series signals or images, these signals possess complex, irregular
structure, which requires novel processing techniques leading to the emerging field of signal processing on graphs~\cite{ShumanNFOV:13,SandryhailaM:14}. Since the structure can be represented by a graph, we call these signals as~\emph{graph signals}. The interest in sampling and recovery of graph signals has increased in the last few years~\cite{Pesenson:08,MarquesSLR:16,ChenVSK:15,ChenVSK:16,puy2017structured,TsitsveroBD:16,RomeroIG:17}. Previous works have however studied sampling strategies on the entire graph in question which can often be expensive both in terms of computational and sample complexity. In this work, we present a structured sampling and recovery framework on product graphs. Product graphs are graphs that are composed of smaller~\emph{graph atoms}; we motivate how this model is a flexible and useful way to model richer data that may be multi-modal in nature~\cite{LeskovecCKFG:10,SandryhailaM:14}.

For example, product graph~\emph{composition} using a product operator is a natural way to model time-varying signals on a sensor network as shown in Figure 1(b). The graph signal formed by the measurements of all the sensors at all the time steps is supported by the graph that is the product of the sensor network graph and the time series graph. The $k^{th}$ measurement of the $n^{th}$ sensor is indexed by the $n^{th}$ node of the $k^{th}$ copy of the sensor network graph.

Multiple types of graph products exist, that is, we can enforce connections across modes in different ways~\cite{weichsel1962kronecker}. In the case of the Cartesian product as in Figure 1(b), the measurement of the $n^{th}$ sensor at the $k^{th}$ time step is related to not only to its neighboring sensors at the $k^{th}$ time step but also to its measurements at the $(k-1)^{th}$ and $(k+1)^{th}$ time steps respectively. Hence, constructing a framework for efficient sampling and recovery on such product graphs is an important step for tasks such as graph signal recovery, compression, and semi-supervised learning on large-scale and multi-modal graphs.
    
In~\cite{ChenVSK:15}, a sampling theory for signals that are bandlimited on graphs was presented.
That is, it was shown that perfect recovery is possible for graph signals bandlimited under the graph Fourier transform. In this paper, we extend this sampling theory by showing how to efficiently sample and recover bandlimited signals on product graphs.

\section{Graphs and Product Graphs}

\begin{figure}
\begin{center} 
\includegraphics[width=31mm]{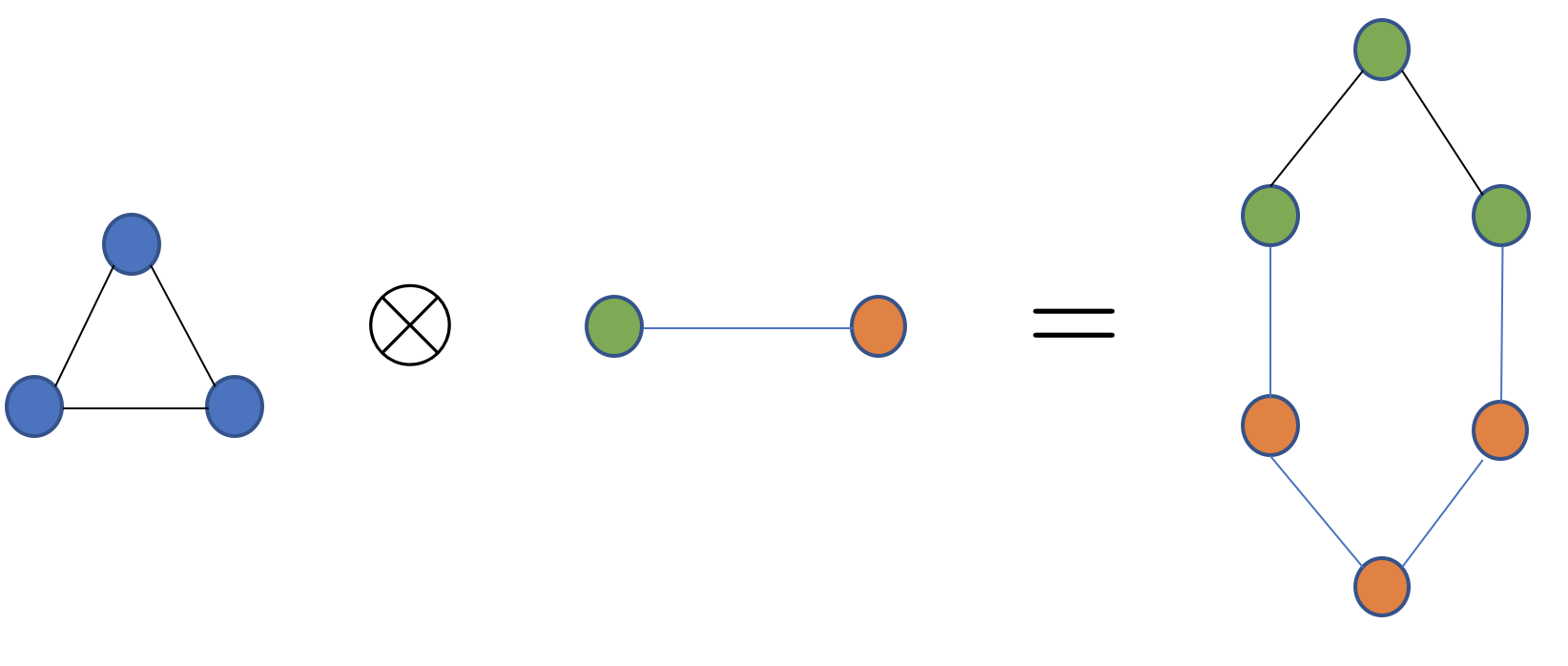}
\hspace{2mm}
\includegraphics[width=50mm]{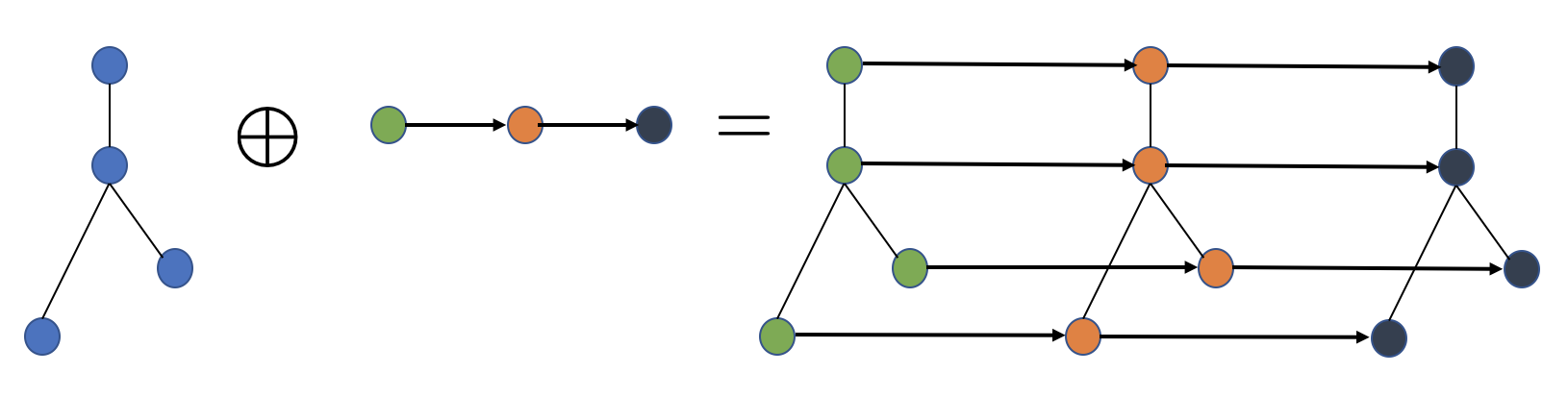} \newline%
(a) \hspace{30mm} (b)
\end{center} 
 \caption{ (a) Under the Kronecker product,  $(u_1,u_2) \sim (v_1,v_2)$ in the product graph if  $u_1\sim v_1$ and $u_2 \sim v_2$.  (b) Under the Cartesian product,  $(u_1,u_2) \sim (v_1,v_2)$ in the product graph if   $u_1 = v_1$ and $u_2 \sim v_2$ or $u_1 \sim v_1$ and $u_2 = v_2$}
\end{figure}

 We consider a  graph $G = (\V,\Adj)$, where $\V = \{v_0,\ldots, v_{N-1}\}$ is the set of nodes
and $\Adj \in \R^{N \times N}$ is the graph shift, or a weighted adjacency matrix.  $\Adj$ Represents the connections of the graph $G$, which can be
either directed or undirected. The
edge weight $w(n \rightarrow m ) = \Adj_{n,m}$ between nodes $v_n$ and $v_m$ is a quantitative expression of the underlying relation between the $n^{th}$ and the $m^{th}$ node, such as a similarity, a dependency, or a communication
pattern. If there exists a non-zero edge weight between $v_n$ and $v_m$, we write $ v_n \sim v_m $.
Once the node order is fixed, the graph signal is  written as a vector
\begin{equation}
\label{eq:graph_signal}
\nonumber
  \x \ = \ \begin{bmatrix}
 x_0, x_1, \ldots, x_{N-1}
\end{bmatrix}^T \in \R^N.
\end{equation}

 Product graphs are graphs whose adjacency matrices are composed using the~\emph{product} (represented by the square symbol $\square$) 
 of the adjacency matrices of smaller~\emph{graph atoms}. Consider two graphs $G_1 = (\V_1, \Adj_1)$ and $G_2 = (\V_2, \Adj_2)$ . The graph product of $G_1$ and $G_2$ is the graph $G = G_1 \square G_2 = (\V , \Adj_1 \square \Adj_2 )$ where $ | \V | = | \V_1 | \cdot | \V_2 | $. The set of nodes $\V$ is the Cartesian product of the sets $\V_1$ and $\V_2$. That is, a node $(u_1,u_2)$ is created for every  $u_1 \in \V_1$ and $u_2 \in \V_2$. 
 


Typically, we use one of the Kronecker graph product ($\otimes$, Figure 1(a)), the Cartesian graph product ($\oplus$, Figure 1(b) or the strong graph product ($\boxtimes$) which is a combination of both the Kronecker and Cartesian product to compose a product graphs. Since the product is associative, one can extend the above formulation to define product graphs constructed from multiple graph-atoms. 

 Digital images reside on rectangular lattices that are Cartesian products of line graphs for rows and columns. We have already seen how the Cartesian product is a natural way to analyze time-varying signals on graphs by enforcing further connections both across the graph in question and the time graph.  A social network with multiple communities can also be represented by the Kronecker graph product of the graph that represents a community structure and the graph that captures the interaction between neighbors. In the context of recommender engines where we have user ratings for different entities at different times, we can view this as a signal lying on the Kronecker product of three graphs, the graph relating the different users, the graph relating the different entities, and the time graph. In the context of multivariate signals on a given graph $\Adj$ where each node has a multidimensional vector associated with it, we can view this as a signal lying on the product graph constructed by the composition of $\Adj$ and the covariance matrix of the multivariate data $\bm{\Sigma}$.
 
   In the following exposition, for clarity and brevity, we only consider the Kronecker product. However, the results and theorems either hold or can easily be extended to both Cartesian and strong products. We also only consider the graph Fourier transform defined for the graph shift matrix $\Adj$ but these results can also be extended for when the graph Fourier transform is defined for the graph Laplacian.   

\section{Graph Signal Processing and the Graph Fourier Transform}

\subsection{Single Graph}

 The spectral decomposition of $\Adj$ is $\Adj=\Vm \Lambda \Vm^{-1}$
where the eigenvectors of $\Adj$ form the columns of matrix $\Vm$~\cite{VetterliKG:12}. We note that $\Vm^{-1} = \Vm^T$ if $\Adj$ is symmetric, that is, the graph $G$ is undirected. $\Lambda\in\R^{N\times N}$ is the diagonal matrix of corresponding
ordered eigenvalues $\lambda_0, \, \ldots, \, \lambda_{N-1}$ of $\Adj$. These eigenvalues represent frequencies on the graph~\cite{SandryhailaM:131}.\\ 

The~\emph{graph Fourier transform} of $\x \in \R^N$ is $ \widehat{\x} = \Vm^{-1} \x$ and the~\emph{inverse graph Fourier transform} is $\x  =  \Vm  \widehat{\x}$.
The vector $\widehat{\x}$ represents the
signal's expansion in the eigenvector basis of the graph shift and describes the frequency content of the graph signal $\x$. The inverse graph Fourier transform reconstructs the graph signal from its frequency content by aggregating graph frequency components weighted by the coefficients of the
signal's graph Fourier transform. 

\subsection{Product Graphs}

 We consider a product graph $G = (\V, \Adj), | \V | = N$, that is constructed from $J$ graph atoms $G_1, G_2, \cdots G_J$, where $G_j= (\V_j, \Adj^{j}), |\V_j | = N_j$, using the Kronecker product where $\prod_{j=1}^J N_j = N$. 
We can write the resulting graph shift matrix of the product graph as 
\begin{equation}
\Adj = \Adj ^{(1)} \otimes \Adj ^{(2)} \otimes \cdots \otimes \Adj ^{(J)} = \otimes_{j =1}^J \Adj ^{(j)}
\end{equation}
We can then write the spectral decomposition of the product graph shift $\Adj$ as
\begin{align}
\Adj &= \Vm  \Lambda  \Um \\ 
\textnormal{where }\Vm &= \Vm^{(1)}  \otimes \Vm^{(2)} \otimes \cdots \otimes \Vm^{(J)} = \otimes_{j =1}^J \Vm^{(j)} \\ 
\Lambda &= \Lambda^{(1)}  \otimes \Lambda^{(2)} \otimes \cdots \otimes \Lambda^{(J)} = \otimes_{j =1}^J \Lambda^{(j)} \\ 
\Um &= \Um^{(1)}  \otimes \Um^{(2)} \otimes \cdots \otimes \Um^{(J)} = \otimes_{j =1}^J \Um^{(j)}
= \Vm^{-1} 
\end{align}

For a given graph atom,  $G_j$, the columns of $\Vm^{(j )}$ and their corresponding frequencies are pairs of the form $(\vm^{(j )}_{i_{(j )}}, \lambda^{(j )}_{i_{(j )}})$. Here, $i_{(j )}$ is an index for the nodes in $G_j$ that varies from $(1,2, \cdots N_j )$ where $N_j  = |  \V_j  |$, the number of nodes in $\G_j $.

As a result, under the Kronecker Product, each of the $N$  basis vectors in $\Vm$ have the form 
\begin{equation}
(  \vm^{(1 )}_{i_{(1)}} \otimes \cdots \otimes \vm^{(j )}_{i_{(j)}} \otimes \cdots \otimes \vm^{(J)}_{i_{(J)}} ,  \lambda^{(1 )}_{i_{(1)}} \times \cdots \times  \lambda ^{(j)}_{i_{(j)}} \times \cdots \times  \lambda^{(J)}_{i_{(J)}} ) \\
\end{equation}
  across all combinations of the indices $ (i_{(1)} , \cdots, i _{(j)} , \cdots , i_{(J)})$. For example, if $\Vm^{(1)} = [\vm^{(1)}_1  | \vm^{(1)}_2]$ and $\Vm^{(2)} = [\vm^{(2)}_1  | \vm^{(2)}_2 |  \vm^{(2)}_3]$,
\begin{align*}
 \Vm^{(1)} \otimes \Vm^{(2)} =  [ \vm^{(1)}_1 \otimes \vm^{(2)}_1 | \vm^{(1)}_1 \otimes \vm^{(2)}_2 | \vm^{(1)}_1 \otimes \vm^{(2)}_3 | \cdots \\ 
 \vm^{(1)}_2 \otimes \vm^{(2)}_1 | \vm^{(1)}_2 \otimes \vm^{(2)}_2 | \vm^{(1)}_2 \otimes \vm^{(2)}_3 ] 
\end{align*}

\section{Sampling Theory: Bandlimited Graph Signals}

In this section, we show how to efficiently sample and recover bandlimited signals on product graphs. We first briefly overview the sampling theory for a single graph before extending the framework to the product graph setting. 

Formally, we can define a bandlimited signal on a graph as follows: 
\begin{defn}
  \label{df:BL}
  A graph signal $\x \in \R^N$ is~\emph{bandlimited} on a graph $\Adj$ when there exists a $K \in
  \{0, 1, \cdots, N-1\}$ such that its graph Fourier transform
  $\widehat{\x}$ satisfies
  \begin{displaymath}
  \widehat{x}_k  =  0 \quad {\rm for~all~}  \quad k \geq K.
\end{displaymath}
Denote this class of graph signals by $\BL_{K}(\Vm)$~\cite{ChenVSK:15}.
\end{defn}

We employ the spectrum-aware setting where we know the support of the signal in the graph Fourier domain. Since we can order the support (eigenvectors) arbitrarily, we refer to these signals as bandlimited signals. 

Suppose that we want to sample exactly $J$ coefficients in a graph signal $\x \in
\R^N$ to produce a sampled signal $\x_\M \in \R^J$ ($M < N$). We then interpolate $\x_{\M}$ to get
$\x' \in \R^N$, which recovers $\x$ either exactly or
approximately. The sampling operator $\Psi$ corresponding to sampling set $\M \subset [n] $ is a linear mapping from
$\R^N$ to $\R^J$, defined as
\begin{equation}
\label{eq:Psi}
 \Psi_{i,j} = 
  \left\{ 
    \begin{array}{rl}
      1, & j = \M_i;\\
      0, & \mbox{otherwise},
  \end{array} \right. 
\end{equation}
and the interpolation operator $\Phi$ is a linear
mapping from $\R^J$ to $\R^N$ 

\begin{eqnarray}
{\rm sampling:}~~&&\x_{\M} =  \Psi \x \in \R^{M},
\nonumber
\\ \nonumber
{\rm interpolation:}~~&&\x' =  {\Phi} \x_{\M} = \Phi \Psi \x  \in \R^{N}.
\end{eqnarray}
Perfect recovery happens for all $\x$ when $\Phi \Psi$ is the identity matrix. This
is not possible in general because ${\rm rank}(\Phi \Psi) \leq M <
N$.

\begin{myThm}
\label{thm:sg}
~\cite{ChenVSK:15}. Let $\Psi$ be the sampling operator to sample $K$ coefficients in $\x \in \BL_{K}(\Um)$ to produce $\x_\M \in \R^K$ and satisfy
  \begin{equation}
    \nonumber
    {\rm rank}( \Psi \Vm_{(K)}) = K.
  \end{equation} 
 Let $\Wm $ be $(\Psi\Vm_{(K)})^{\dagger}$.  Perfect recovery is then achieved by setting 
\begin{equation}
\Phi = \Vm_{(K)} \Wm 
\end{equation}
such that $ x = \Phi \x_\M$.  \\ 
In addition,  $\x_\M$ is a graph signal associated with the graph shift 
\begin{equation}
  \Adj_\M = \Wm^{-1} \Lambda_{(K)} \Wm \in \R^{K \times K}.
\end{equation}
whose graph Fourier transform is $\Wm$. 
\end{myThm}

Theorem~\ref{thm:sg} shows how to sample and perfectly recover bandlimited graph signals on graphs. In addition, we see that the sampled graph signal lies on a~\emph{sampled} graph shift $\Adj_\M$. Since the bandwidth of $\x$ is $K$, the first $K$ coefficients in the frequency domain are $\widehat{\x}_{(K)} = \widehat{\x}_\M$, and the other $N-K$ coefficients are $\widehat{\x}_{(-K)} = 0$. In other words, the frequency contents are equivalent for the original graph signal $\x$ and the sampled graph signal $\x_\M$ after operation of the   graph Fourier transform that corresponds to the graph they are associated with.
The sample size $J$ should be no smaller than the
bandwidth $K$. We also note that  at least one set of $K$ linearly-independent rows in $\Vm_{(K)}$ always exists.

\subsection{Sampling Theory: Product Graph}

We now consider the product graph $G$  that is composed using the Kronecker product over  $J$ graphs  $\{ \G^{(1)}, \cdots, \G^{(J)}\} $.

As before, we have a bandlimited graph signal $ \x \in \BL_K(\Vm)$ that is associated with the product graph $\Adj$ and a sampling operator $\Psi$ such that the sampled signal $\xm  = \Psi \x$ is acquired by applying the sampling operator $\Psi$. We showed in Theorem~\ref{thm:sg} that a sufficient condition to perfectly recover the sampled  bandlimited signal $\xm = \Psi \x $ where $ \x \in \BL_K(\Vm)$ is that 
\begin{equation}
\rank(\Psi \Vm_{(K)} ) = K
\end{equation}

It is straightforward to sample the product graph using the framework constructed in the previous section for a single graph by  using the composed graph-shift  $\Adj$ as a whole. Instead, in this section, we look to exploit the structure  of the product graph under the Kronecker product composition when we sample the graph. We note here that we are free to order the eigenvectors of $\Vm$ arbitrarily. 

We have seen that we can write any given column vector $\vm$ of $\Vm$  as a particular combination of $J$ column vectors from each of the $\Vm ^{(j)}$:
\begin{equation}
\vm = \vm^{(1 )}_{i_{(1)}} \otimes \cdots \otimes \vm^{(j )}_{i_{(j )}} \otimes \cdots \otimes \vm^{(J )}_{i_{(J)}} = \bigotimes_{j = 1}^J  \vm^{(j )}_{i_{(j )}}
\end{equation}
where $\vm^{(j )}_{i_{(j )}}$ is a column of $\Vm ^{(j)}$ indexed by $i_{(j )}$.

Given some subset of $K$ columns of $\Vm$ over which the signal is bandlimited, we can accordingly re-order the columns in each of $\Vm ^{(j)}$ such that 
\begin{equation}
\Vm_{(K)} \subset \Vm_{R_{1}}^{(1)} \otimes \cdots  \Vm_{R_{j}} ^{(j)} \otimes \cdots  \Vm_{R_{J}}^{(J)}  = \bigotimes_{j = 1}^J  \Vm_{R_{j}} ^{(j)} = \Vm_{S}.
\end{equation}

 $\Vm_{ R_{j}}^{(j)}$ corresponds to the top $ R_{j} $ columns of $\Vm ^{(j)}$ and $S = \prod_{j = 1}^J R_j$. We note that $K \leq S  \leq K^J$.
In addition, any signal that is in $\BL_K (\Vm)$ is also in $\BL_{S} ( \bigotimes_{l = 1}^J  \Vm_{R_j} ^{(j)}).$

\begin{myThm}
\label{thm:spg}
Let us consider the sampling scheme where we sample $R_{j}$ nodes from each of the sub-graphs $\G ^{(j)}$  using the sampling operator $\Psi ^{(j)}$. 

Using Theorem~\ref{thm:sg}, for each of the $J$ graph atoms, we can construct appropriate sampling ($\Psi ^{(j)}$) and interpolation ($\Phi ^{(j)})$ operators corresponding to the subset of columns $R_j$ in $\Vm ^{(j)}$ such that for any $\x ^{(j)} \in \BL_{R_j} (\Vm ^{(j)})$, we can sample and perfectly recover such that $\x ^{(j)} = \Phi ^{(j)} ( \Psi^{(j)} \x ^{(j)} ) = \Phi ^{(j)} \x_\M ^{(j)}$. In addition, $\x_\M ^{(j)}$ is associated with a sampled graph whose graph shift is $\Adj ^{(j)}_\M$. \\ 
 We now sample $S$ nodes in the product graph corresponding to all combinations of the sampled nodes in the graph atoms.
That is, we construct the sampling operator $\Psi$ to sample $ S = \prod_{j = 1}^J R_{j} $ nodes in the product graph such that $\x_\M = \Psi \x$
\begin{equation}
\label{eq:prodsamp}
\Psi = \bigotimes_{j = 1}^J \Psi ^{(j)}
\end{equation}
Further, the corresponding interpolation operator $\Phi$ over the product graph is   
\begin{equation}
\label{eq:prodrec}
\Phi = \bigotimes_{j = 1}^J \Phi^{(j)} 
\end{equation}
As a result, $\Psi$ and $\Phi$ enable perfect recovery such that for any bandlimited graph signal $\x \in \BL_K (\Vm)$ on the product graph,
$\x = \Phi \x_\M = \Phi \Psi \x$.
\end{myThm}
\begin{proof}
Full proof omitted due to lack of space. We can write
\begin{align}
\Psi \Vm_{(K)} = \bigotimes_{j = 1}^J \Psi^{(j)} \Vm_{R_{(j)}}^{(j)}  \text{ and } \rank(\Psi \Vm_{(K)}) = \prod_{j=1}^J R_{(l)}  
\end{align}
We recognize that in order to satisfy the condition $ \rank(\Psi \Vm_{(K)} ) \geq K$, it is sufficient to ensure that for each of the graph atoms, $\rank(\Psi ^{(j)} \Vm_{R_j} ^{(j)} ) \geq   R_j$. 
\qedhere
\end{proof}
In addition, the sampled graph signal $\x_\M$ lies on a sampled product graph. Particularly the sampled product graph can be decomposed as the Kronecker product of the sampled graph for the individual sub-graphs. That is,
\begin{equation}
\Adj_{\mathcal{M}} = \bigotimes_{j = 1}^J \Adj_{\mathcal{M}}^{(j)}
\end{equation}

 \begin{figure}
   \begin{center}
      \includegraphics[width= 1.05\columnwidth]{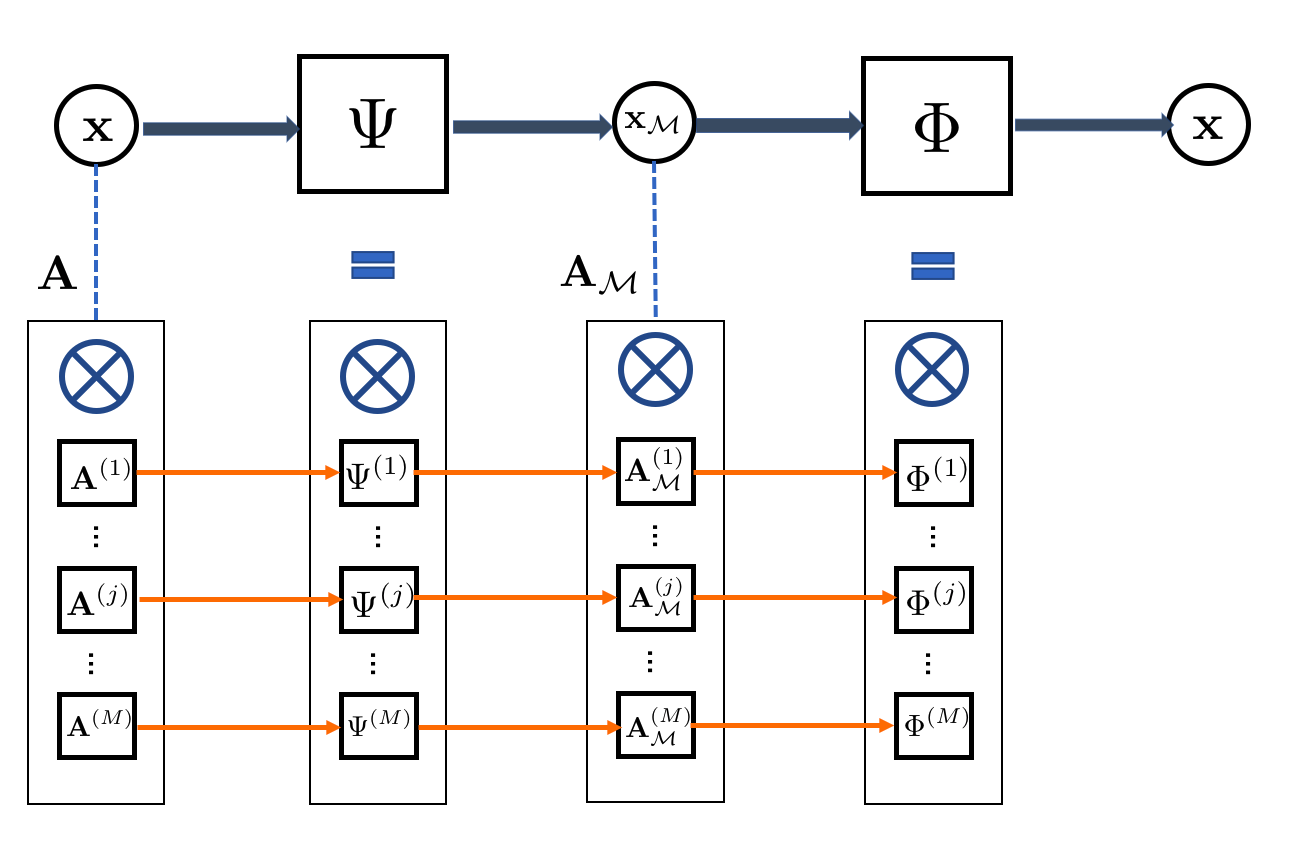}
   \end{center}
   \caption{\label{fig:flow} As shown in Theorem~\ref{thm:spg}, we can construct an admissible sampling operator and corresponding interpolation operator by composing sampling and interpolation operators defined respectively on each of the graph atoms. Further, the sampled graph signal lies on a sampled product graph }
 \end{figure}

The sampling and recovery framework for product graphs based the decomposition of the sampling and interpolating operators presented in Theorem~\ref{thm:spg} is illustrated in Figure~\ref{fig:flow}.

\begin{figure*}[!htb]
  \includegraphics[width=\textwidth,height=5.5cm]{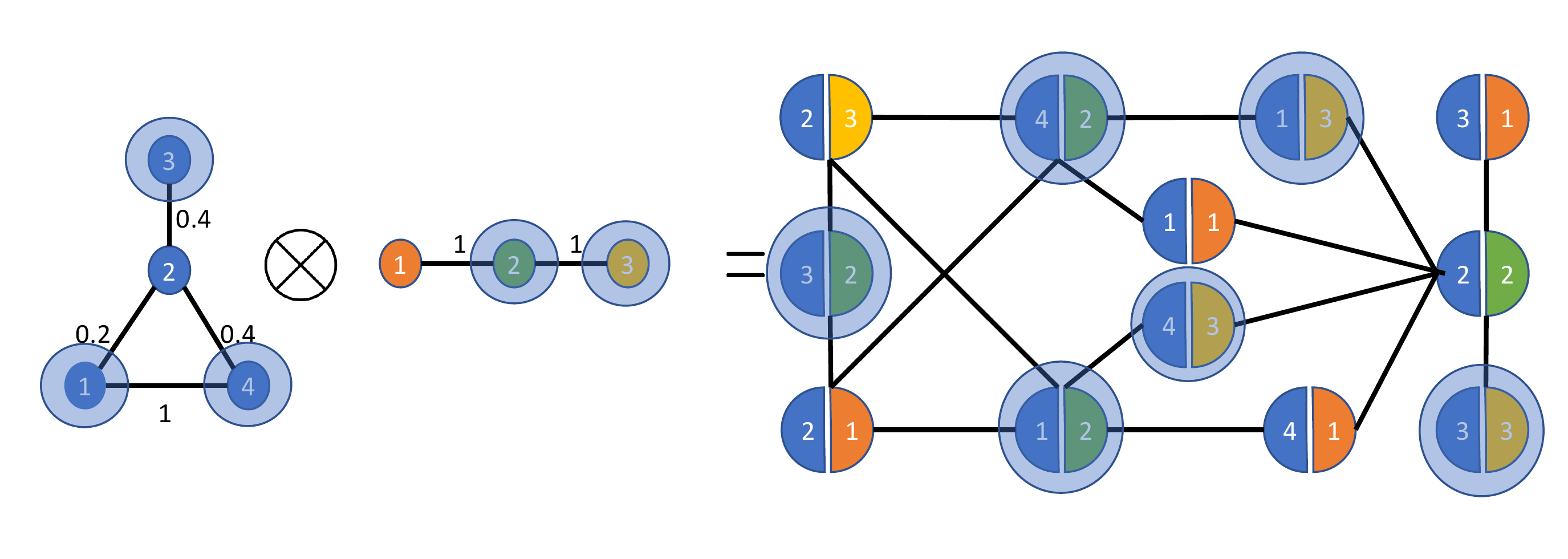}
  \caption{ In Section IV-B,  we consider sampling and recovering a signal $ \x \in \BL_K(\Vm)$ with $K=3$. The top $K=3$ columns of the ordered GFT basis $\Vm$ corresponds to the pairs $(1,1),(4,3)$, and $(3,3)$ of the graph atoms respectively. We choose a sample set consisting of nodes (1,3,4) on the $\Adj_1$  and a sampling set (2,3) on $\Adj_2$. The sample sets are marked by the transparent blue circle in the above figure. We then sample nodes corresponding to all combinations of the sampling sets on the product graph. That is, we sample 6 nodes corresponding to the following pairs $\{ (1,2), (1,3) , (3,2),(3,3),(4,2),(4,3) \} $ We can appropriately construct an interpolation operator from the interpolation operators corresponding to the chosen sampling sets on the graph atoms $\Adj_1$ and $\Adj_2$ such that we can ensure perfect recovery for any bandlimited signal $ \x \in \BL_K(\Vm)$}
  \label{fig:toy}
\end{figure*}

\subsection{Toy Example}

In this section, we study a toy example that further illustrates Theorem~\ref{thm:spg}. As shown in Figure~\ref{fig:toy}, consider a graph $\Adj = \Adj_1 \otimes \Adj_2$ and a bandlimited signal $ \x \in \BL_K(\Vm)$ with K=3. The top $K=3$ columns of the ordered GFT basis $\Vm$ corresponds to the pairs $(1,1),(4,3)$, and $(3,3)$ from the graph atoms respectively. As a result, we can set $R_{1} = \{ 1,3,4\} $ and $R_2 = R_{1} = \{ 1,3\}$ such that $\Vm_{(K)} \subset \Vm_{R_{1}}^{(1)} \otimes \Vm_{R_{2}}^{(2)}$. We can then compose sampling and interpolation operators using Theorem~\ref{thm:sg} for each of the two graphs:
\[ \Psi^{(1)} =
\begin{bmatrix}
   0  &   1   &  0   &  0 \\
     0  &   0  &   1   &  0 \\
     1  &   0  &   0   &  0 \\
\end{bmatrix}
\hspace{2mm} \Phi^{(1)} = 
\begin{bmatrix}
   0 & 0 & 1 \\   
    1 & 0 & 0 \\
   0 & 1 & 0 \\
    0.69  &  2.21 &  -1 \\
\end{bmatrix} 
\]

\[ \Psi^{(2)} =
\begin{bmatrix}
 0   &  1  &   0 \\
     0  &   0  &   1 \\

\end{bmatrix}
\hspace{15mm} \Phi^{(2)} = 
\begin{bmatrix}
  0 & 1 \\ 
    1 & 0 \\ 
      0 & 1 \\ 
\end{bmatrix} 
\]

We then compose the sampling and interpolation operators as in Theorem~\ref{thm:spg} as $\Psi = \Psi^{(1)} \otimes \Psi^{(2)}$ and $\Phi =\Phi^{(1)} \otimes \Phi^{(2)}$. We then sample $ |R_1 | |R_2 | = 6 $ nodes in the product graph as $\xm = \Psi \x$ corresponding to combinations of the chosen sampling sets for each of the graph atoms as illustrated in Figure~\ref{fig:toy}. We then see that we can sample and perfectly reconstruct any bandlimited signal $ \x \in \BL_K(\Vm)$ by verifying that $\Phi \xm = \x $.

\subsection{Discussions}
\emph{Sample Complexity:} We have seen that we need at least $K$ samples in order to perfectly recover a bandlimited graph signal $\x \in BL_K(\Vm)$ in the single graph setting. In the product graph sampling framework prescribed above, we need atleast $S$ samples of the graph signal on the product graph where $K \leq S \leq K^J$. Hence, in the worst case, we need $K^J$ samples to ensure perfect recovery. \\
Smooth signals on graphs are approximately bandlimited under a fixed frequency ordering~\cite{ChenVSK:16}. We can show that under the Cartesian product, we only need $S \leq K + J $ samples to perfectly recover and sample a smooth signal that is in $BL_K(\Vm)$ which is nearly optimal. 

\emph{Computational Complexity}: We note that we do not need to process the whole product graph $\Adj$ or compute its spectral decomposition (GFT basis) which is of complexity $O(N^3)$ and is often computationally prohibitive for large graphs. Instead we can construct sampling and interpolation operators on the product graph using only the spectral decompositions of its graph atoms $\Adj^{(j)}$ that are of size $O(poly(N^{\frac{1}{J}}))$. 
We choose  $R_{j}$ nodes from each of the graphs $G^{(j)}$ and  sample  $ S = \prod_{j = 1}^J R_j $ nodes in the product graph such that each sampled node in the product graph correspond to some combination of the sampled nodes in the graph atoms. Hence, we effectively only need to do~\emph{choose} $\sum_{j = 1}^J R_j$ nodes over the graph atoms. In contrast, in the single graph setting, we need to choose atleast $K$ nodes, where in general $ K = O(S)$.

\emph{Kronecker Graphs:}  In~\cite{LeskovecCKFG:10}, a generative model that can effectively model the structure of many large real-world networks was presented by recursively applying the Kronecker product on a base graph that can be estimated efficiently. We can consequently leverage our framework to sample graph signals that are supported on a large real-world networks with a substantial reduction in the sample and computational complexity. 

\emph{Miscellaneous:} We can extend the above sampling procedure on a product graph under the noisy sample acquisition setting by finding the optimal sampling operator. In addition, we can process multi-band signals by sampling optimally on a product graph by constructing filter banks analogously to~\cite{ChenVSK:15}. 

\section{Conclusion}
 In this paper, a framework for efficient sampling and recovery of bandlimited signals on product graphs was presented. Particularly, we showed that by exploiting the structure of a product graph and designing appropriate sampling and recovery operators on the graph atoms that the product graph is composed of, we achieve significant savings in sample and computational complexity.

\bibliographystyle{IEEEbib}
\bibliography{bibl}

\end{document}